\documentclass[pra,twocolumn,showpacs,a4paper,superscriptaddress]{revtex4}

\usepackage{graphicx}
\usepackage{amsmath}
\usepackage{amsfonts}
\usepackage{amssymb}

\newcommand{\beq}{\begin{equation}}
\newcommand{\eeq}{\end{equation}}
\newcommand{\beqr}{\begin{eqnarray}}
\newcommand{\eeqr}{\end{eqnarray}}
\newcommand{\lb}[1]{\label{#1}}
\newcommand{\ct}[1]{\cite{#1}}

\begin{document}

\title{Self-cooling of a movable mirror to the ground state using radiation pressure}

\author{A. Dantan}
\affiliation{QUANTOP, Danish National Research Foundation Center for
Quantum Optics, Department of Physics and Astronomy, University of
Aarhus, DK-8000 \AA rhus C., Denmark}
\author{C. Genes}
\affiliation{Dipartimento di Fisica, Universit\`a di Camerino,
via Madonna delle Carceri, I-62032 Camerino (MC), Italy}
\author{D. Vitali}
\affiliation{Dipartimento di Fisica, Universit\`a di Camerino, via
Madonna delle Carceri, I-62032 Camerino (MC), Italy}
\author{M. Pinard}
\affiliation{Laboratoire Kastler Brossel, Universit\'{e} Pierre et
Marie Curie, Case 74, 4 place Jussieu, 75252 Paris Cedex 05, France}

\date{\today}

\begin{abstract}
We show that one can cool a micro-mechanical oscillator to its
quantum ground state using radiation pressure in an appropriately
detuned cavity (self-cooling). From a simple theory based on
Heisenberg-Langevin equations we find that optimal self-cooling
occurs in the good cavity regime, when the cavity bandwidth is
smaller than the mechanical frequency, but still larger than the
effective mechanical damping. In this case the intracavity field and
the vibrational mechanical mode coherently exchange their
fluctuations. We also present dynamical calculations which show how
to access the mirror final temperature from the fluctuations of the
field reflected by the cavity.
\end{abstract}


\maketitle

Cooling of mechanical resonators both at the micro- and at the
macro-level, is an important technical challenge in various fields
of physics, such as ultra-high precision measurements \cite{schwab},
and detection of gravitational waves \cite{GW}. It is also a
prerequisite for any possible use of optomechanical systems for
quantum information processing \cite{prltelep,prl07}. Active noise
control techniques have been proposed to reduce their thermal noise
and bring the oscillator motion to its ground state \cite{grassia}.
Recently, various experiments have demonstrated significant cooling
of the vibrational mode coupled to an optical cavity
\cite{cohadon99,karrai04,gigan06,arcizet06,arcizet06b,bouwm,vahalacool,mavalvala,rugar,harris}.
The experiments of
Refs.~\cite{gigan06,arcizet06b,vahalacool,mavalvala} in particular
have adopted the so-called \emph{back-action} \cite{brag}, or
\emph{self-cooling} scheme, in which the radiation pressure of an
appropriately detuned cavity interacting with the mechanical
oscillator, is used. In all these experiments, however, the
resulting equilibrium state of the oscillator is classical, because
the new mean excitation number is still much larger than one.
Therefore it is important to establish the fundamental limits of
self-cooling and if it is able to cool a mechanical degree of
freedom down to its quantum ground state. We will describe the
system dynamics in terms of quantum Langevin equations (QLE) for the
Heisenberg operators of the system and we will show that ground
state self-cooling is possible in the good cavity regime, i.e. when
the mechanical resonance frequency is larger than the cavity
bandwidth, and provided the cavity detuning and bandwidth are
appropriately chosen so as to maximize the scattering of noisy
photons into the cavity mode.

We consider a driven optical cavity coupled by radiation pressure to
a micromechanical oscillator. The typical experimental configuration
is a Fabry-Perot cavity with one mirror much lighter than the other
(see e.g.
\cite{karrai04,gigan06,arcizet06,arcizet06b,bouwm,harris}), but our
treatment applies to other optomechanical systems, such as the
silica toroidal microcavity of Refs.~\cite{vahala1,vahalacool}.
Radiation pressure typically excites several mechanical degrees of
freedom of the system with different resonant frequencies. However,
we consider the detection in a narrow frequency bandwidth,
containing a single mechanical resonance peak. In this case the
oscillator dynamics can be well approximated by that of a single
harmonic oscillator with dimensionless position $q(t)$, and momentum
$p(t)$, $([q,p]=2i)$, frequency $\Omega_{\mathrm{m}}$, mass $M$, and
damping $\Gamma$. The coupled dynamics of the mirror with the
intracavity field mode $a\left(t\right)$ in the frame rotating at
the frequency of the driving laser $\omega_L$, is described by the
QLE \begin{eqnarray}
\dot{q}&=&\Omega_m p\\
\dot{p}&=&-\Omega_m q-\Gamma p+G\sqrt{2}a^{\dagger}a+\xi\\
\dot{a}&=&-\left(\kappa+i\Delta_c\right)a+iGaq/\sqrt{2}+
\sqrt{2\kappa}a^{\rm in}, \label{eq:InOut1}
\end{eqnarray} where $G=(\omega_c/L)\sqrt{\hbar/M\Omega_m}$ is the opto-mechanical
coupling constant, $\omega_c$ and $L$ being the cavity resonance
frequency and length, respectively. $a^{\rm in}\left(t\right)$ is
the input field, satisfying $\langle a^{\rm in}(t)a^{\rm
in\dagger}(t')\rangle-\left|\overline a^{\rm
in}\right|^2=\delta(t-t')$, with $\overline a^{\rm in}$ the incident
mean field. $\kappa$ is the damping rate of the cavity mode,
$\Delta_c=\omega_c-\omega_L$ the cavity detuning and $\xi$ the noise
operator accounting for the mirror Brownian motion at thermal
equilibrium at temperature $T$ \ct{note0}.\\

\textit{Steady state analysis -} In steady state the mean
intracavity field $\overline a$ is given by $\overline a =
\sqrt{2\kappa}\overline a^{\rm in}/(\kappa+i \Delta)$, where
$\Delta$ is the mean detuning of the cavity, given by $\Delta =
\Delta_c-\Delta_{nl}=\Delta_c - G^2|\overline a|^2/\Omega_m$. These
two coupled equations give a third-order relation between $\overline
a$ and $\Delta$ which leads to the well-known bistable behavior of a
cavity with a movable mirror \cite{Dorsel83}. The stability
condition of the system can be written as
$\kappa^2+\Delta^2+2\Delta\Delta_{nl}> 0$.

\begin{figure}[htb]
\centerline{\includegraphics[width=8cm,clip=]{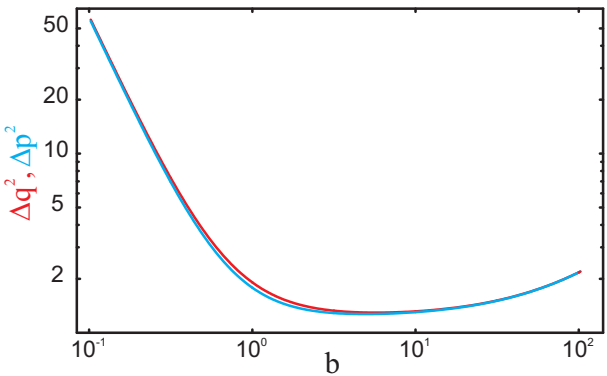}}
\caption{Normalized variances versus $b$, for a detuning $\varphi=b$
$Q=10^4$, $n_T^i=10^2$, $\varphi_{nl}=0.1$. The final excitation
number is $n_T^f\sim 0.15$ for $b\sim 5$.} \label{fig1}
\end{figure}

If we then consider the linearized fluctuations of the various
operators around the steady state, we get linearized QLE which can
be solved by moving to the frequency domain. The Fourier transform
of the position fluctuations can then be simply expressed as the sum
of a thermal noise term and a radiation pressure term \beqr
q[\Omega]=\tilde{\chi}[\Omega]\left(F_R[\Omega]+F_T[\Omega]\right),\eeqr
the response to the noise terms being given by an effective
susceptibility \beqr
\tilde{\chi}[\Omega]^{-1}=\chi[\Omega]^{-1}-M\Omega_m^2\frac{2\Delta_{nl}\Delta
}{\kappa^2D[\Omega]}\eeqr with $\chi[\Omega]=
M\Omega_m^2\left[1-\Omega^2/\Omega_m^2-i\Gamma\Omega/\Omega_m^2\right]$
and $ D[\Omega]=(1-i\Omega/\kappa)^2+(\Delta/\kappa)^2$. The thermal
noise force is simply $F_T[\Omega]=M\Omega_m\xi[\Omega]$ and the
radiation pressure force \cite{note1}  \beqr\nonumber
F_R[\Omega]=\frac{\sqrt{2}M\Omega_m^{\frac{3}{2}}\Delta_{nl}^{\frac{1}{2}}}{\kappa^{\frac{3}{2}}D[\Omega]}\left[(\kappa-i\Omega)x^{\rm
in}[\Omega]+\Delta y^{in}[\Omega] \right]\eeqr with $x^{\rm
in}=a^{\rm in}+a^{\rm in\dagger}$ and $y^{\rm in}=i(a^{\rm
in\dagger}-a^{\rm in})$. An exact expression for the mirror
variances is then obtained by integrating the contributions of these
two forces to the noise spectrum $(\omega=\Omega/\Omega_m)$
\beqr\Delta q^2=\int \frac{d\omega}{2\pi} S_q(\omega),
\hspace{0.3cm}\Delta p^2=\int \frac{d\omega}{2\pi}\omega^2
S_q(\omega)\lb{varex0}\eeqr where we have used the fact
that$p=\dot{q}/\Omega_m$ and where the noise spectrum is given by
\begin{widetext} \beqr S_q(\omega)= \left[\frac{2\omega
\coth\left(\frac{\hbar\Omega_m \omega}{2k_BT}\right)
}{Q}+4\varphi_{nl}\frac{1+\varphi^2+b^2
\omega^2}{(1-b^2+\varphi^2)^2+4b^2\omega^2}\right]\frac{|(1-i b
\omega)^2+\varphi^2|^2}{|[(1-i b
\omega)^2+\varphi^2][1-\omega^2-i\omega/Q]-2\varphi\varphi_{nl}|^2},\lb{varex}
\eeqr\end{widetext} with $b=\Omega_m/\kappa$,
$\varphi=\Delta/\kappa$, $\varphi_{nl}=\Delta_{nl}/\kappa$ and $
Q=\Omega_m/\Gamma$. These two exact expressions for the variances
coincide with Eq. 5 of Ref.~\cite{genes}. Cooling to the ground
state implies reaching $\Delta q^2= 1$ simultaneously with $\Delta
p^2= 1$. Fig.~\ref{fig1} shows both variances as a function of $b$,
for a cavity detuning equal to one bandwidth ($\varphi=b$), which
shows that, starting from a mean thermal excitation number
$n_T^i=\left(\exp\{\hbar \Omega_m/k_BT\}-1\right)^{-1}$ equal to
$10^2$, it is possible to achieve a lower
than unity final excitation number state in the \textit{good} cavity limit: $b>1$, i.e. $\Omega_m>\kappa$.\\

\textit{Physical interpretation in the adiabatic limit -} To get
more insight into the cooling mechanism we will now derive an
analytical expression for the frequency integrals of (\ref{varex0}).
The latter may be considerably simplified by noting that, for high-Q
cavities and under certain conditions that we will detail, the
mirror response is peaked around the natural mechanical resonance
frequency. One may then express the effective susceptibility as
$\tilde{\chi}[\Omega]^{-1}\simeq
M\tilde{\Omega}^2[1-\Omega^2/\tilde{\Omega}^2-i/\tilde{Q}]$ with an
effective quality factor $\tilde{Q}=\tilde{\Omega}/\tilde{\Gamma}$
and effective resonance frequency and relaxation rate \beqr
\tilde{\Omega}&=&\Omega_m\left[1-2\varphi\varphi_{nl}\;\textrm{Re}[D[\Omega_m]^{-1}\right]^{1/2}\\
\tilde{\Gamma}&=&\Gamma\left[1+2\varphi\varphi_{nl}Q\;\textrm{Im}[D[\Omega_m]^{-1}\right]\label{gammaeff}\eeqr
The effect of the frequency-shift in the resonance is typically
negligible, so that one may assume $\tilde{\Omega}\simeq \Omega_m$.
The relaxation rate of the mirror is on the contrary strongly
affected by the presence of the cavity field and, depending on the
sign of the cavity detuning, may either be enhanced or reduced,
resulting in either cooling or heating. As pointed out in
\cite{wilsonrae,epl,marquardt}, the most interesting regime for
self-cooling is when the effective damping rate $\tilde{\Gamma}$ is
strongly enhanced, but still less than the cavity bandwidth and when
the effective mechanical $\tilde{Q}$ is still larger than one,
$\Gamma\ll\tilde{\Gamma}\ll\kappa$. In this case, the effective
susceptibility is still peaked around $\omega=1$ and we can well
approximate the smoothly varying function of $\omega$ inside the
square brackets by taking its value at $\omega=1$. One then gets an
analytical expression for the variance \beq\lb{varapp} \Delta
q^2\simeq \frac{\Gamma}{\tilde{\Gamma}}
\left[2n_T^i+1+2\varphi_{nl}Q\frac{1+b^2+\varphi^2}{(1-b^2+\varphi^2)^2+4b^2}\right]\eeq
Eq.~(\ref{varapp}) is the basic equation for our analysis of the
quantum limits of self-cooling, and it is possible to see that it
coincides with the results of \cite{marquardt,wilsonrae}, which were
obtained with a different method. Ref.~\cite{genes} instead gives
slightly more general expressions for the two variances, which,
however, reduce to that of Eq.~(\ref{varapp}) when $\Omega_m$ is
large enough, i.e, $\Omega_m
> \tilde{\Gamma}$.

\begin{figure}[htb]
\centerline{\includegraphics[width=8cm,clip=]{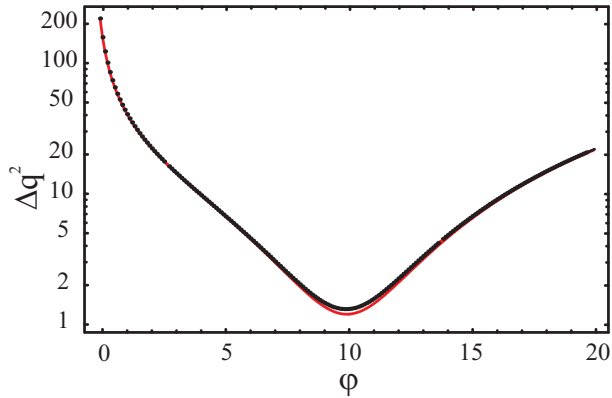}}
\caption{Normalized variance versus cavity detuning [dashed: exact
result of Eq.~(\ref{varex}), plain: approximate result of
Eq.~(\ref{varapp})]. Parameters: $Q=10^4$, $n_T^i=10^2$,
$\varphi_{nl}=0.1$, $b=10$.} \label{varphi}
\end{figure}

Depending on the respective frequencies involved, one can adopt
different strategies to cool the mirror down to the ground state. In
particular, the cavity detuning has to be well-chosen in order to
optimize the self-cooling. Fig.~\ref{varphi} shows the variation of
the variance with the cavity detuning, as calculated from the exact
solution of Eq.~(\ref{varex}) and from the harmonic approximation
(\ref{varapp}). The analytical result is in good agreement with the
numerical result and we will use it as a basis for the cooling
optimization discussion.\\

In order to
interpret Eq.~(\ref{varapp}) we rewrite it as \beq\lb{varapp1}
\Delta
 q^2=\frac{1+2n_T^i}{1+f\varphi_{nl}Q}+\frac{f\varphi_{nl}Q}{1+f\varphi_{nl}
Q}\frac{1+b^2+\varphi^2}{2\varphi b},\eeq or, equivalently,
\beq\Delta q^2=(1-\eta)\;\Delta q_T^2+\eta\; \Delta
q_R^2.\lb{varapp2}\eeq $\Delta q_T^2=1+2n_T^i$ and $\Delta
q_R^2=(1+b^2+\varphi^2)/2\varphi b$ represent the thermal
noise-induced and the cavity field-induced contributions,
respectively, to the position fluctuations of the mirror.
\beq\eta=\frac{f\varphi_{nl}Q}{1+f\varphi_{nl}Q}\eeq is the weight
between those two quantities, with \beq f=\frac{4\varphi
b}{(1-b^2+\varphi^2)^2+4b^2}.\eeq The physical interpretation of
Eq.~(\ref{varapp2}) is that the cavity field and the mirror
vibrational mode coherently exchange their fluctuations during the
interaction, all the more so that the quantity $\eta$ is close to
unity. The thermal excess noise is then replaced by the cavity field
fluctuations, which are those of the vacuum, hence resulting in
decreasing the effective temperature of the mirror. The quantity
$\eta$ naturally appears as a quantum state transfer efficiency
between the field and the vibrational mode, whereas the product
$\varphi_{nl}Q$ can be interpreted as a coherent coupling strength,
quite similarly to the \textit{cooperativity} parameter, which
appears in quantum state transfer schemes between optical fields and
atomic ensembles \ct{dantan1,dantan2,gorshkov}. Since the mirror is
initially in a noisy thermal state, the cavity field acts as a
thermal noise ``eater" during the interaction, and allows to reach a
much lower effective temperature for the mirror in steady state.
This situation is also quite similar to cavity assisted
Doppler-cooling of atoms \cite{cavitycooling}, for which the
temperature of the atoms is decreased by enhancing the scattering of
photons into the cavity mode.

If one starts with a very large thermal excess noise,
Eq.~(\ref{varapp1}) clearly shows that the thermal noise
contribution will be suppressed by a factor $f\varphi_{nl}Q$, which
is typically large. Minimizing the mirror final temperature is then
equivalent to maximizing the transfer efficiency $\eta$, and, for a
given value of $\varphi_{nl}Q$, maximizing $f$ with respect to
$\varphi$. This gives an optimal cavity detuning equal to
\beq\lb{phit}\varphi^*=[(b^2-1+2\sqrt{1+b^2+b^4})/3]^{\frac{1}{2}}.\eeq
The most favorable conditions to reach the quantum ground state of
the mechanical oscillator then occur for a cavity bandwidth
\emph{smaller} than the mechanical resonance. Indeed, assuming $b\gg
1$, one has $\varphi^*\simeq b$, $f^*\simeq 1$ and for a
sufficiently high Q-factor ($\varphi_{nl} Q\gg 2n_T^i+1$), the
resulting mirror fluctuations are then mostly given by the radiation
pressure: \beq (\Delta q^2)^*\simeq \Delta q_R^2\simeq
1+\frac{1}{2b^2}\eeq and the normalized variance tends to unity for
$b\gg 1$. It is therefore possible to reach the mechanical ground
state using self-cooling in the \emph{good} cavity limit.

However, there exists an optimal bandwidth that minimizes the mirror
final temperature. Indeed, as one decreases the cavity bandwidth
(for a fixed value of the mechanical resonance frequency), the
effective response time of the mirror becomes comparable to that of
the intracavity field and the previous adiabatic approximation is no
longer valid. One can also show that this situation is detrimental
to the noise transfer between the field and the mirror
\cite{dantan1,dantan2}. As mentioned previously and as can be seen
in Fig.~\ref{fig1}, the adiabatic approximation
breaks down when $\varphi_{nl}\Omega_m\gtrsim 2\kappa$.\\

\begin{figure}[htb]
\centerline{\includegraphics[width=7.0cm,clip=]{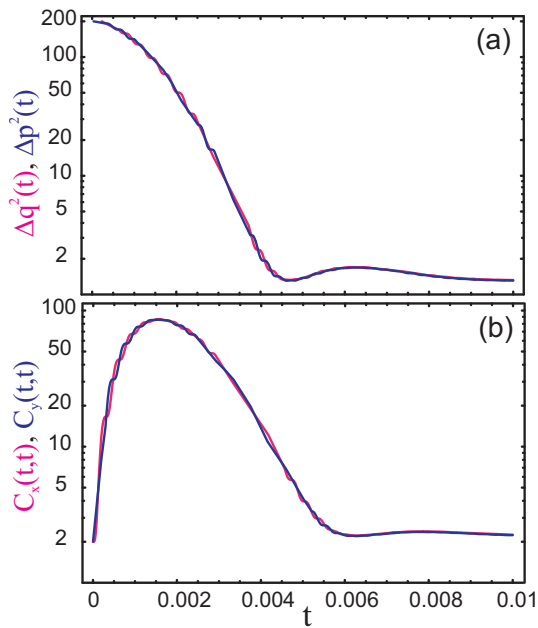}}
\caption{(a) Temporal evolution of the mirror position and momentum
normalized variances when the cooling field is on from $t=0$
($\Gamma^{-1}$ units). (b) Temporal evolution of the outgoing field
variances $C_{x,y}(t,t)$ (in $\kappa$ units). Parameters:
$\kappa=10^3\Gamma$, $Q=10^4$, $n_T^i=10^2$, $\varphi_{nl}=0.1$,
$\varphi=b=10$.} \label{dyn}
\end{figure}
\textit{Dynamical situation -} We now address the dynamical behavior
of the coupled field-mirror system. Under good self-cooling
conditions one expect the mirror initial thermal noise to be mainly
transferred to the cavity field during the cooling phase, and hence
to the field leaking out of the cavity. Numerically solving the
linearized Langevin equations in time allows to compute the two-time
correlation functions of $q(t)$ and $p(t)$, as well as those of the
outgoing field quadratures, $x^{\rm out}$ and $y^{\rm out}$, where
$a^{\rm out}=\sqrt{2\kappa}a-a^{\rm in}$. To simplify, we assume a
step-like injection of a coherent state cooling field into the
cavity at $t=0$ and start with an occupation number $n_T^i=10^2$ for
the mirror~\cite{note2}. The resulting mirror variances at time $t$
are represented in Fig.~\ref{dyn}, as well as the temporal evolution
of the equal-time variances of the outgoing field $C_{x,y}(t,t)$,
obtained from exact numerical calculations. It appears clearly that,
during the cooling phase, thermal noise is taken from the mirror and
transferred to the cavity field. The amount of thermal noise taken
from the mirror can be deduced by measuring the outgoing field
fluctuations. Indeed, for $\Omega_m=\Delta$ and in the adiabatic
limit $\Gamma\ll\tilde{\Gamma}\ll\kappa$, it can be shown that the
outgoing field correlation functions are of the form
\cite{epl,note4} \beqr C_{x,y}(t,t')
=2\eta\tilde{\Gamma}e^{-\tilde{\Gamma}(t+t')}+n_T^{i}\eta(1-\eta)\tilde{\Gamma}e^{-|t-t'|}\eeqr
Following the method developed in Ref.~\cite{dantan3}, one can
perform a homodyne detection of the outgoing field fluctuations with
a temporally matched local oscillator, $\mathcal{E}_{LO}(t)\sim
e^{-\tilde{\Gamma} t}$ \ct{note3}. The measured quantity is then
$\Delta x_m^2=1+\eta (n_T^i-n_T^f)+\eta(1-\eta)n_T^i,$ where
$n_T^{i,f}$ are the initial and final thermal excitation numbers.
When the
cooling is optimal ($\eta\sim 1$), one indeed measures the change in temperature of the mirror: $\Delta x_m^2\simeq 1+n_T^i-n_T^f$.\\

We have presented a general theory for the self-cooling of a
mechanical oscillator to the ground state, which allows for deriving
analytical results for the mirror final temperature and provides a
simple interpretation of the cavity mediated self-cooling process.
When the cavity is suitably detuned the initial thermal noise of the
mirror vibrational mode is essentially transferred to the cavity
field mode. This noise exchange reflects in the field leaking out of
the cavity, providing useful information on the mirror temperature.


\begin{thebibliography}{99}

\bibitem{schwab} M.D. LaHaye, \textit{et al.}, Science \textbf{304}, 74
(2004).

\bibitem{GW} C. Bradaschia \textit{et al.}, Nucl. Instrum. Methods Phys.
Res. A \textbf{289}, 518 (1990); A. Abramovici \textit{et al.},
Science \textbf{256}, 325 (1992); P. Fritschel, Proc. SPIE
\textbf{4856}, 282 (2002).

\bibitem{prltelep} S. Mancini, \textit{et al.}, Phys. Rev. Lett. \textbf{90}
, 137901 (2003).

\bibitem{prl07} D. Vitali \textit{et al.}, Phys. Rev. Lett. \textbf{98},
030405 (2007).

\bibitem{grassia} F. Grassia \textit{et al.}, Eur. Phys. J. D
\textbf{8}, 101 (2000); J.M. Courty, A. Heidmann, and M. Pinard,
Eur. Phys. J. D \textbf{17}, 399 (2001).

\bibitem{cohadon99} P.F. Cohadon \textit{et al.}, Phys. Rev. Lett. \textbf{
83}, 3174 (1999).

\bibitem{karrai04} C.H. Metzger and K. Karrai, Nature (London), \textbf{432}, 1002
(2004).

\bibitem{gigan06} S. Gigan \textit{et al.}, Nature (London) \textbf{444}, 67
(2006).

\bibitem{arcizet06} O. Arcizet, \textit{et al.}, Phys. Rev. Lett. \textbf{97}%
, 133601 (2006).

\bibitem{arcizet06b} O. Arcizet, \textit{et al.}, Nature (London) \textbf{444}%
, 71 (2006).

\bibitem{bouwm} D. Kleckner and D. Bouwmeester, Nature (London) \textbf{444}%
, 75 (2006).

\bibitem{vahalacool} A. Schliesser \textit{et al.}, Phys. Rev. Lett. \textbf{%
97} 243905 (2006).

\bibitem{mavalvala} T. Corbitt \textit{et al.}, Phys. Rev. Lett. \textbf{98}%
, 150802 (2007).

\bibitem{rugar} M. Poggio \textit{et al.}, e-print cond-mat/0702446.

\bibitem{harris} J.G.E. Harris \textit{et al.}, Rev. Sci. Instrum. \textbf{%
78}, 013107 (2007).

\bibitem{brag} V.B. Braginsky \textit{et al.}, Phys. Lett. A \textbf{287},
331 (2001).

\bibitem{vahala1} T.J. Kippenberg \textit{et al.}, Phys. Rev. Lett. \textbf{%
95} 033901 (2005).

\bibitem{note0} $\langle \xi(t)\xi(t')\rangle=(\Gamma/\pi\Omega_m)\int d\Omega
e^{-i\Omega(t-t')}\Omega[\coth(\frac{\hbar\Omega}{2k_BT})+1]$.

\bibitem{Dorsel83} A. Dorsel \textit{et al.}, Phys. Rev. Lett.
\textbf{51}, 1550 (1983).

\bibitem{note1} We assume $\overline a$ real by choosing the input field phase.

\bibitem{genes}C. Genes \textit{et al.}, e-print quant-ph/0705.1728.

\bibitem{wilsonrae}I. Wilson-Rae \textit{et al.}, e-print cond-mat/0702113.

\bibitem{epl} M. Pinard \textit{et al.},
%A. Dantan, D. Vitali, O. Arcizet, T. Briant and A. Heidmann,
Europhys. Lett. \textbf{72}, 747 (2005).

\bibitem{marquardt}F. Marquardt \textit{et al.}, e-print cond-mat/0701416.

\bibitem{dantan1} A. Dantan and M. Pinard, Phys. Rev. A \textbf{69}, 043810 (2004).

\bibitem{dantan2} A. Dantan, A. Bramati, and M. Pinard, Europhys. Lett. \textbf{67}, 881 (2004).

\bibitem{gorshkov} A. Gorshkov \textit{et al.}, Phys. Rev. Lett. {\bf 98}, 123601
(2007).

\bibitem{cavitycooling} V. Vuleti\'{c} and S. Chu, Phys. Rev.
Lett. {\bf 84}, 3787 (2000).

\bibitem{note2} This analysis can
straightforwardly be extended to arbitrary non-stationary
interaction, based on the atomic ensemble/light
analogy~\ct{dantan3}.

\bibitem{dantan3} A. Dantan, J. Cviklinski, M. Pinard, and P. Grangier, Phys. Rev. A
{\bf 73}, 032338 (2006).

\bibitem{note4} from which the vacuum contribution has been subtracted, ie $C_{u}(t,t')=\langle
u^{\rm out}(t)u^{\rm out}(t')\rangle-\delta(t-t')$ ($u=x,y$).

\bibitem{note3} This corresponds to measuring the fluctuations of
$a_m=\int dt \mathcal{E}_{LO}(t)a_{\rm out}(t)$.

\end{thebibliography}
\end{document}